\documentclass{article}

\usepackage{booktabs}
\usepackage{graphics}
\usepackage{graphicx}
\usepackage{listings}
\usepackage{nameref}
\usepackage{algorithm}
\usepackage{algpseudocode}
\algrenewcommand\algorithmicrequire{\textbf{Input:}}
\algrenewcommand\algorithmicensure{\textbf{Output:}}
\usepackage{bm}
\usepackage{amsmath}
\usepackage{amssymb}
\usepackage{caption} 
\usepackage{multirow}
\usepackage{color}
\usepackage[table]{xcolor}

\usepackage{arxiv}

\usepackage[utf8]{inputenc} % allow utf-8 input
\usepackage[T1]{fontenc}    % use 8-bit T1 fonts
\usepackage{hyperref}       % hyperlinks
\usepackage{url}            % simple URL typesetting
\usepackage{booktabs}       % professional-quality tables
\usepackage{amsfonts}       % blackboard math symbols
\usepackage{nicefrac}       % compact symbols for 1/2, etc.
\usepackage{microtype}      % microtypography
\usepackage{lipsum}
\usepackage{graphicx}
\graphicspath{ {./images/} }

\title{PhaBOX: A web server for identifying and characterizing phage contigs in metagenomic data}

\author{
 Jiayu Shang \\
  Dept. of Electrical Engineering\\
  City University of Hong Kong\\
  Kowloon, Hong Kong SAR, China\\
  \texttt{jyshang2-c@my.cityu.edu.hk} \\
  \And
 Cheng Peng \\
  Dept. of Electrical Engineering\\
  City University of Hong Kong\\
  Kowloon, Hong Kong SAR, China\\
  \texttt{cpeng29-c@my.cityu.edu.hk} \\
  \And
 Herui Liao \\
  Dept. of Electrical Engineering\\
  City University of Hong Kong\\
  Kowloon, Hong Kong SAR, China\\
  \texttt{heruiliao2-c@my.cityu.edu.hk} \\
  \And
 Xubo Tang \\
  Dept. of Electrical Engineering\\
  City University of Hong Kong\\
  Kowloon, Hong Kong SAR, China\\
  \texttt{xubotang2-c@my.cityu.edu.hk} \\
  \And
 Yanni Sun \\
  Dept. of Electrical Engineering\\
  City University of Hong Kong\\
  Kowloon, Hong Kong SAR, China\\
  \texttt{yannisun@cityu.edu.hk} \\
}

\begin{document}

\maketitle
\begin{abstract}
\textbf{Motivation:} There is accumulating evidence showing the important roles of bacteriophages (phages) in regulating the structure and functions of the microbiome. However, lacking an easy-to-use and integrated phage analysis software hampers microbiome-related research from incorporating phages in the analysis.  \\
\textbf{Results:} In this work, we developed a web server, PhaBOX, which can comprehensively identify and analyze phage contigs in metagenomic data. It supports integrated phage analysis, including phage contig identification from the metagenomic assembly, lifestyle prediction, taxonomic classification, and host prediction. Instead of treating the algorithms as a black box, PhaBOX also supports visualization of the essential features for making predictions. The web server is designed with a user-friendly graphical interface that enables both informatics-trained and non-specialist users to analyze phages in microbiome data with ease. \\
\textbf{Availability:} The web server of PhaBOX is available via:  \href{https://phage.ee.cityu.edu.hk}{https://phage.ee.cityu.edu.hk}. The source code of PhaBOX is available at: \href{https://github.com/KennthShang/PhaBOX}{https://github.com/KennthShang/PhaBOX}. \\
\textbf{Contact:} \href{yannisun@cityu.edu.hk}{yannisun@cityu.edu.hk}\\
\end{abstract}

\section{Introduction}
\label{sec:intro}
As viruses that infect bacteria, bacteriophages (phages) are the most widely distributed and abundant biological entities in the biosphere \cite{cobian2016viruses}. With an estimated population of more than $10^{31}$ particles \cite{mushegian2020there}, phages play an important role in modulating microbial system dynamics by lysing bacteria \cite{fernandez2018phage}. Recently, accumulating studies show that phages have an important impact on multiple applications, such as disease diagnostics \cite{roth2021developing}, phage therapy \cite{petrovic2020safety}, and on the food industry \cite{cristobal2021phages}.

Although there are tools available for different tasks such as phage taxonomic classification and host prediction, these tools are often published as open-source codes. When the users want to characterize phages in metagenomic data, they need to install several tools and parse the intermediate files from different methods. However, installing some of these softwares requires informatics training and takes substantial computational resources. For example, some open-source codes are not accompanied with detailed user manuals, making the installation tedious and error-prone. Thus, it is preferred to have an integrated web server that can conduct comprehensive phage analysis while sparing the users the time and trouble of installing multiple open-source tools. 
%Because phages ubiquitously exist in many different ecosystems, such as soil and marine samples, we develop this web server to accommodate the needs of users from different fields and help advance the field of phage discovery in different types of ecosystems.

In this work, we present a toolbox for phage analysis (PhaBOX), a comprehensive web service for phage identification, lifestyle prediction, taxonomic classification, and host prediction. In PhaBOX, we optimize the integrated tools to improve the running speed and efficiency. In addition, PhaBOX can visualize the essential features that are important for making the final predictions, such as the similarity-based relationships between the input sequence and other phages, predicted proteins on the input sequences, and protein homology. PhaBOX can take either metagenomic assemblies or whole genome sequencing assemblies as inputs. All the predictions and intermediate results are provided for downstream analysis.

\begin{figure*}[h!]
    \centering
    \includegraphics[width=0.95\linewidth]{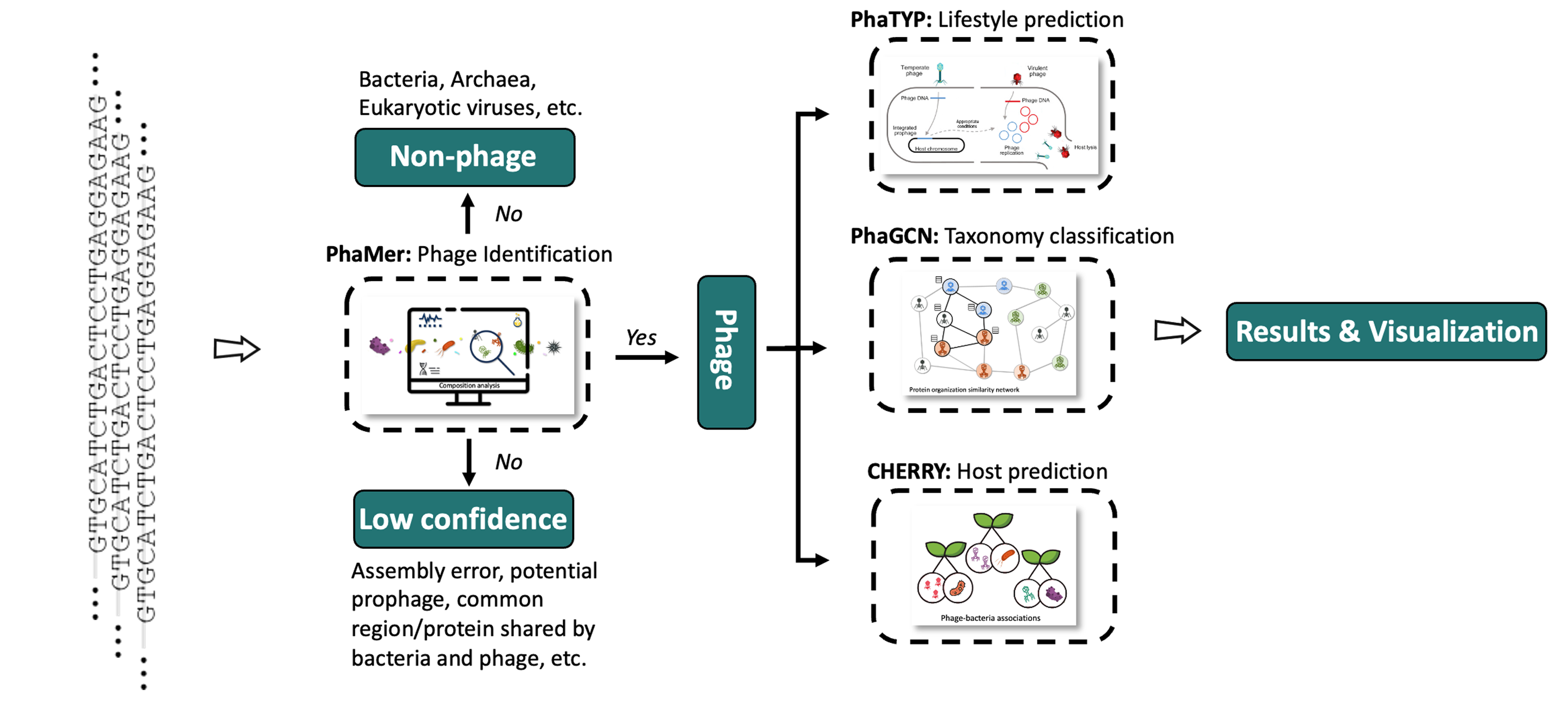}
    \vspace{-0.2cm}
    \caption{The pipelines of PhaBOX. The input of PhaBOX is the FASTA file containing assembled contigs. Then, PhaMer \cite{shang2022accurate} is applied for phage identification. Only the contigs predicted as phages will be used for lifestyle prediction (PhaTYP \cite{shang2023phatyp}), taxonomic classification (PhaGCN \cite{shang2021bacteriophage}), and host prediction (CHERRY \cite{shang2022cherry}). Finally, we provided the predictions and visualization for users on the result page.}
    \label{fig:model}
\end{figure*}

\begin{figure*}[h!]
    \centering
    \includegraphics[width=0.95\linewidth]{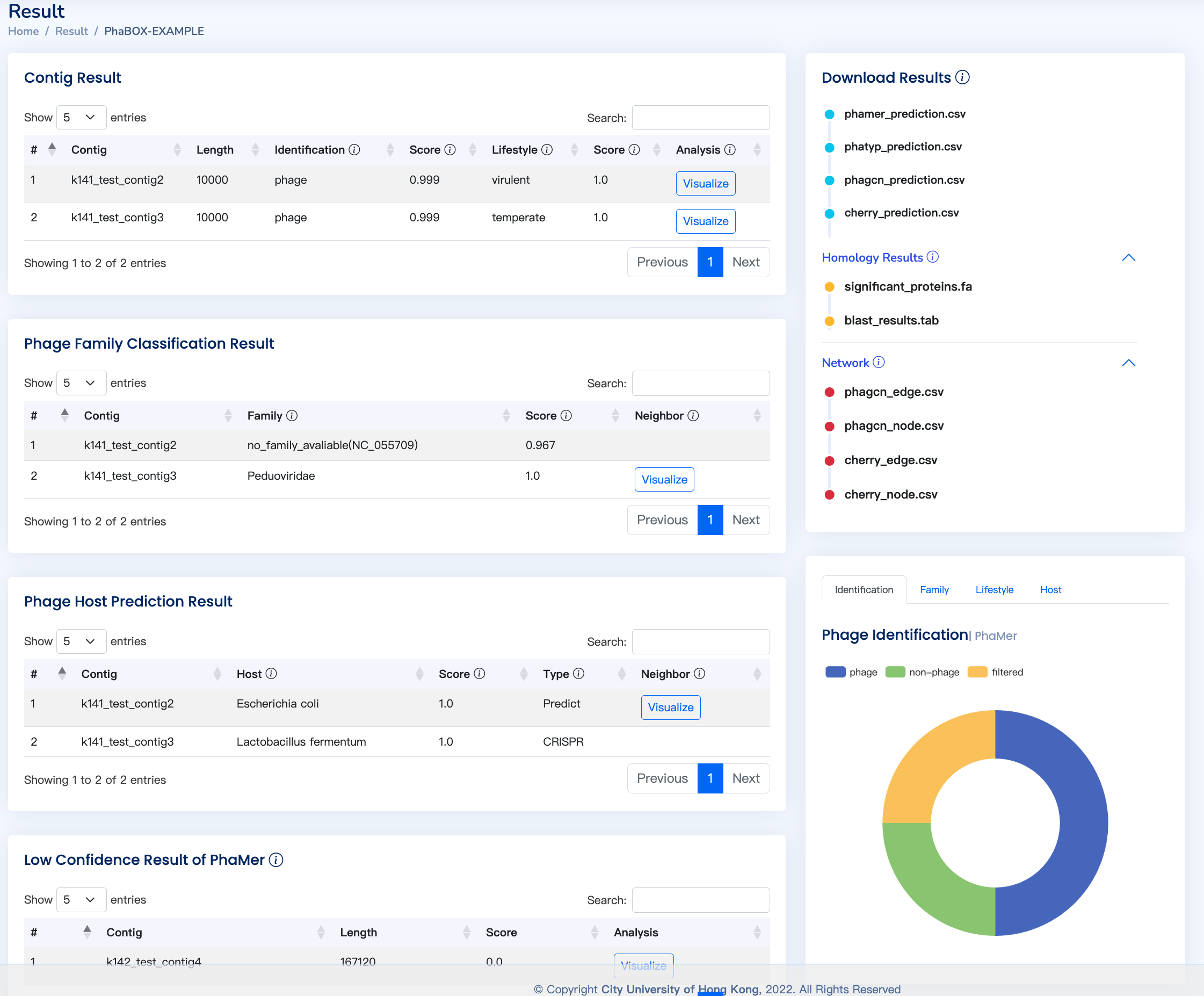}
    \vspace{-0.2cm}
    \caption{An example of the result page. The visualization can be accessed by clicking the visualize button. All the results can be downloaded from the top-right download panel. The meaning of the head will be shown when moving the cursor onto the icon of the header.}
    \label{fig:result}
\end{figure*}

\subsection{Overview}

PhaBOX is a web server developed with Python/R, providing integrative identification and characterization for phage contigs in metagenomic data. Our algorithms behind PhaBOX were peer-reviewed and published, including PhaMer \cite{shang2022accurate} for phage identification, PhaTYP \cite{shang2023phatyp} for lifestyle prediction, PhaGCN \cite{shang2021bacteriophage} for taxonomic classification, and CHERRY \cite{shang2022cherry} for host prediction. All these tools combined the strength of alignment-based strategies and deep learning models to learn different sequence-based features, including protein organizations, sequence homology, and protein-protein associations. Our methods outperform the available programs in each task based on our rigorous tests on highly diverged phages, short contigs, mock metagenomic data, and real metagenomic data. For example, according to a third-party view \cite{tang2022phage}, the earlier version of CHERRY, named HostG \cite{shang2021predicting} has the best performance on predicting the phage-host relationship on the genus level. In CHERRY, we not only further improved the accuracy but also supported host prediction at the species level.

PhaBOX provides integrated phage identification and analysis on one website. It has a modular design and thus supports two modes of running. The default mode allows users to conduct end-to-end analysis including all the supported functions. The other mode allows users to only run the modules they need, e.g. predicting the hosts for recently sequenced phages. As shown in Fig. \ref{fig:model}, the input of PhaBOX is the FASTA file containing assembled contigs from metagenomic data or whole genome sequencing data. Then, PhaMer is applied for phage contig identification. By default, the identified phage contigs will be used as inputs for all modules including lifestyle prediction, taxonomic classification, and host prediction. For each module, the homology search is carried out against the NCBI RefSeq database. We updated the database and optimized the program to save computing resources. PhaBOX has a running time of approximately 50 seconds per 8kbp contig under its default mode. It is \textasciitilde70\% of the total time of running each program individually. We also provided a multi-threads local version of PhaBOX for users who want to analyze phages in a large amount of metagenomic data.

\begin{figure*}[h!]
    \centering
    \includegraphics[width=0.95\linewidth]{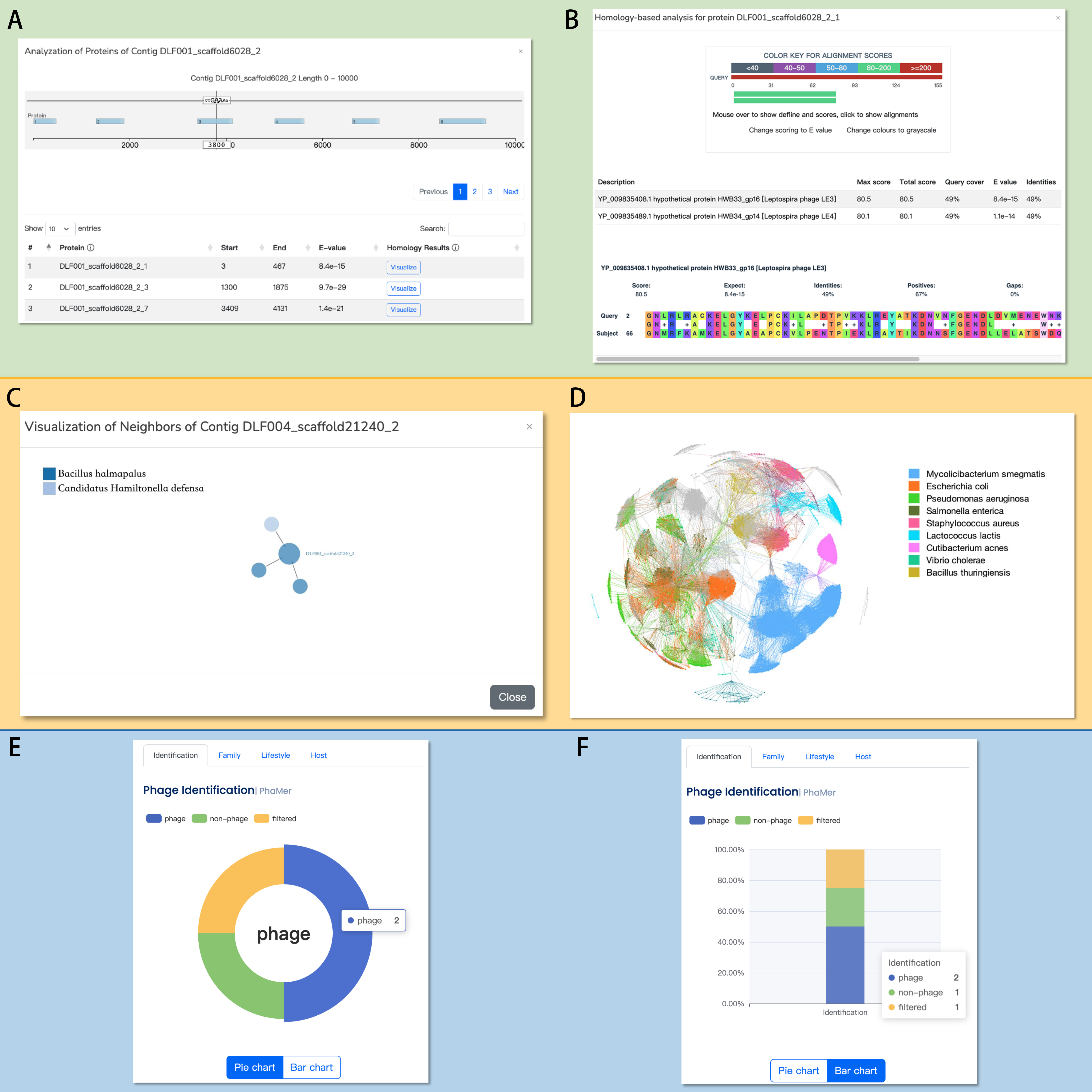}
    \caption{The visualization of the results. (A) The visualization of the homology search on the input contigs. The position of the translated proteins with high alignments score will be shown, and the detailed alignment results (B) can be accessed by clicking the visualize button. (C) We will show the one-step neighbors on the sequence-similarity graph to reveal the relationships between the input contigs and the phages/bacteria in the database. The complete sequence-similarity graph is provided in the ``Network'' panel in Fig. \ref{fig:result}. They are in the standard format for most network-based visualization tools. One example using Gephi to visualize the graph is shown (D). (E) The pie graph of the prediction result. (F) The stacked bar graph is provided as an alternative option.}
    \label{fig:visual}
\end{figure*}

In addition to achieving faster end-to-end phage analysis, PhaBOX provides detailed visualizations of the essential features behind the methodology, which is not available in the previous works. Fig. \ref{fig:result} and Fig. \ref{fig:visual} demonstrate the visualization of essential components, such as similarity-based relationships between contigs and other phages, predicted proteins on the contigs, and protein homology, which provide evidence to support the generation of predictions.

In the following sections, we will showcase how to use PhaBOX for identifying and characterizing phage contigs in metagenomic data. We will demonstrate the utility of PhaBOX and detail the visualization functions provided on the result page.

\section{Case study}
In this case study, we demonstrate that PhaBOX can simplify and speed up phage analysis in human microbiome data. The microbiome data are sequenced from fecal samples of 145 Chinese individuals, including 71 Type 2 diabetes (T2D) patients and 74 controls \cite{qin2012metagenome}. We downloaded the assembled contigs provided by the authors and uploaded the FASTA files to the PhaBOX server. To maintain high-quality predictions, we set the minimum length of contigs as 10kbp as suggested in \cite{qin2012metagenome}. In total, we obtained 129,138 contigs in 145 samples. We used PhaBOX to identify and characterize the phages from the gut metagenomic samples.

\subsection{Example of the result page}

One example result page is shown in Fig. \ref{fig:result}. There are a total of six modules on the result page, including the tab for downloading results,  pie graphs of the prediction, contig results, phage family classification results, phage host prediction results, and low confidence results. The meaning of each module is listed below.

\paragraph{The module for downloading results} As shown in the top-right of Fig. \ref{fig:result}, this module contains all the predictions and intermediate files for users to download. The standard outputs of PhaBOX are four CSV files containing the results of identified phages, their taxonomy, lifestyles, and hosts. PhaBOX also provides intermediate files of the essential features, including the homology of phage-related proteins output by PhaMer/PhaTYP and knowledge network output by PhaGCN/CHERRY. The homology files containing phage-related proteins in FASTA format and BLAST alignment results in CSV format. The knowledge network contains several types of interactions between metagenomic assemblies and phages/bacteria in the database, such as gene-sharing information, CRIPSRs, and sequence similarity. The network files are in standard input formats for most network-based visualization tools, such as Gephi \cite{bastian2009gephi} and Cytoscape \cite{shannon2003cytoscape}.

\paragraph{Graphs of prediction}  In the PhaBOX web server, pie graphs are employed to visualize the final prediction results and stacked bar graphs are provided as alternative options. There are eight charts in total, including the results of the percentage of identified phages, taxonomic composition, lifestyle composition, and host prediction. As shown in Fig. \ref{fig:visual} E/F, the pie graph and stacked bar graph show the percentage of identified phages.

\paragraph{Contig results} As shown in Fig. \ref{fig:result}, the top-left table contains all the identified phages and their lifestyles predicted by PhaMer and PhaTYP. The visualization of the homology search on the input contigs can be accessed by clicking the ``Visualize'' buttons. Then, Fig. \ref{fig:visual} A will be presented to show the position of the translated proteins with their alignments score. Finally, each protein's detailed alignment results can be accessed as shown in Fig. \ref{fig:visual} B.

\paragraph{Phage family classification results and host prediction results} As shown in Fig. \ref{fig:result}, the middle-left tables contain the family-level taxonomic classification results predicted by PhaGCN and host prediction results predicted by CHERRY. Because both methods utilize sequence similarity and gene organization as features to construct phage-phage and phage-bacterium relationship networks, we provide the visualization of these networks. However, the complete similarity network is too large to be visualized on a web page. Thus, we only show the one-step neighborhood of each contig as shown in Fig. \ref{fig:model} F. The visualized network is interactive, and users can drag the nodes to check their accessions. As mentioned in \textit{download results} module, the complete graph can be conveniently generated using our provided file and any of the commonly used network visualization tools such as Gephi. An example of a visualized complete graph by Gephi is shown in Fig. \ref{fig:model} G.

\paragraph{Low confidence results} As shown in Fig. \ref{fig:result}, the bottom-left table contains all contigs that have alignments with the database, but the confidence scores given by the model are too low to be predicted as phages. There are two possible reasons for the low confidence. First, there may exist assembly errors in the contigs. Second, the contigs contain regions that are shared by phage and bacteria. Thus, we provide the corresponding results and visualization for the user who wants to further analyze these sequences.

\subsection{Outputs of PhaBOX}
We uploaded all the T2D contigs to PhaBOX and analyzed the phage contigs. In addition, we recorded the running time of using PhaBOX and the total time of running each program separately. All the methods are run on Intel\textsuperscript{\textregistered} Xeon\textsuperscript{\textregistered} Gold 6258R CPU with 36 threads. The total time for PhaBOX to analyze 129,138 contigs in 145 samples is 216 minutes, saving 40\% of the running time (328 minutes) of running each program separately.

Then, we summarized the prediction for each sample. There are 4,851 phage contigs identified from the 145 samples. They belong to 18 families and infect 211 kinds of host species. PhaBOX provides multiple files for different functions. The summary file contains the comprehensive analysis for each contig (\textit{prediction\_summary.csv}). Table \ref{tab:table1} shows an example of the summarized prediction. More detailed information about the prediction file, such as the confidence of the prediction, can be found in the supplementary files.

\section{Conclusion}
Because phages ubiquitously exist in many different ecosystems, such as soil and marine samples, we develop this web server to accommodate the needs of users from different fields. It is expected that an easy-to-use web server can help advance the field of phage discovery in different types of ecosystems. The integration of phage identification, lifestyle prediction, taxonomic classification, and host prediction of our platform provides not only comprehensive analysis for metagenomic assemblies but also detailed  visualizations for users. The case study demonstrates that PhaBOX can facilitate users with fast and convenient phage characterization in metagenomic data. Future improvements include adding more functions, such as novel protein annotation and protein structural analysis. We will also upgrade the hardware of our server to provide a faster prediction.

%%%%%%%%%%%%%%%%%%%%%%%%%%%%%%%%%%%%%%%%%%%%%%%%%%%%%%%%%%%%%%%%%%%%%%%%%%%%%%%%%%%%%
%
%     please remove the " % " symbol from \centerline{\includegraphics{fig01.eps}}
%     as it may ignore the figures.
%
%%%%%%%%%%%%%%%%%%%%%%%%%%%%%%%%%%%%%%%%%%%%%%%%%%%%%%%%%%%%%%%%%%%%%%%%%%%%%%%%%%%%%%

\begin{table*}[h!]
\centering
{ 
\begin{tabular}{p{4cm}p{1.2cm}p{1.2cm}p{2.5cm}p{3cm}}
\hline
\textbf{Name}          & \textbf{Length}           & \textbf{Lifestyle} & \textbf{Taxonomy}         & \textbf{Host}                       \\ \hline
DLF001\_scaffold16878\_1 & 12,659 & Virulent  & Drexlerviridae   & Bacteroides fragilis       \\
DLF003\_scaffold56673\_2 & 51,717 & Temperate & Casjensviridae   & Bacillus halmapalus        \\
DLM016\_scaffold64767\_1 & 15,261 & Virulent  & Salasmaviridae   & Lactobacillus plantarum    \\
NLM021\_scaffold58236\_5 & 38,364 & Temperate & Straboviridae    & Lactobacillus gasseri      \\
NLF002\_scaffold2199\_6  & 28,963 & Temperate & Straboviridae    & Clostridioides   difficile \\
NLF005\_scaffold21492\_3 & 14,868 & Temperate & Ackermannviridae & Streptococcus mutans       \\ \hline
\end{tabular}
\caption{Example summary file output by PhaBOX. }
\label{tab:table1}
}
\end{table*}

\section*{Method and materials}

The architecture of the PhageBOX server consists of two major components: a client web interface and a server backend. The client web interface is responsible for submitting the tasks and displaying the output. It was implemented by JS, CSS, jQuery, Bootstrap, and their extension packages. Specifically, the sequence similarity was visualized by BlasterJS, the protein sequence viewer was presented using pViz, and the topological graph structure was drawn using Plotly in R. The server backend is responsible for interacting with users through the web interface, handling users’ input, and executing the whole prediction process. The prediction pipeline contains four functional modules including phage identification, lifestyle prediction, taxonomic classification, and host prediction. In the phage identification and lifestyle prediction tasks, we adopt the state-of-the-art language model, Transformer \cite{vaswani2017attention}, to automatically learn abstract patterns from the ``language'' of phages. In the taxonomic classification and host prediction tasks, we construct a knowledge graph by integrating multiple protein and DNA-based sequence features. Then the graph convolutional neural network is applied to utilize features from both labeled and unlabeled samples. All of the methods are comprehensively benchmarked with state-of-the-art tools on multiple datasets, including RefSeq dataset, low similarity dataset, metagenomic dataset and etc. The former interface was implemented by the fast and lightweight Python-based Flask framework and the extension Python packages. The server backend employs a lite SQL database that stores and updates the job information and status. The scheduling method also allows the architecture to be added to add new computational facilities to meet the increasing demand for predicting ever-accumulating genome-scale data. More detailed information can be found under the ``Home'' tab on the PhaBOX web page.

\section*{Funding}
City University of Hong Kong (Project 9678241, 9440274 and 7005453) and the Hong Kong Innovation and Technology Commission (InnoHK Project CIMDA).

\bibliographystyle{unsrt}  
\bibliography{references}

\end{document}